# Topologically robust programmable logic arrays using light and matter skyrmions

Ruofu Liu[1], An Aloysius Wang[1], Yunqi Zhang[1], Yuxi Cai[1], Yihan Liu[2], Zhenglin Li[1], Yifei Ma[1], Zimo Zhao[1], Runchen Zhang[1], Zhi-Kai Pong[1], Stephen M. Morris[1] and Chao He[1]*

[1]Department of Engineering Science, University of Oxford, Parks Road, Oxford, OX1 3PJ, UK
[2]Tsinghua University, Haidian District, Beijing 100084, China
*Corresponding author: chao.he@eng.ox.ac.uk

## Abstract

Photonic computing offers a low-power, high-bandwidth paradigm for information processing; however, the analogue nature of conventional architectures means that intrinsic noise and fabrication imperfections greatly impact performance, thereby severely limiting scalability. Recent work on optical skyrmions offers a route to overcoming these limitations by exploiting perturbation-resilient topological invariants assigned to the optical field for computation. Crucially, owing to its relative novelty, an architectural perspective on integrating individual components that manipulate topological charge into a functional system remains an important open goal. In this paper, we take concrete steps toward system-level design by introducing a platform-independent architecture for skyrmion-based logic, built around a modular library of topologically robust optical primitives, including generators, converters, registers, and adders. This framework enables the synthesis and arithmetic manipulation of topological numbers within a unified programmable architecture. We then experimentally validate this approach using multichannel arrays, demonstrating accurate charge readout and high robustness against alignment errors and environmental noise. These results provide a scalable foundation for topologically robust programmable logic arrays, paving the way for compact and integrated photonic processing circuits.

## Introduction

Topological textures, including optical skyrmions[1–4] and related structures such as merons[5–7], are spatially varying fields characterised by a quantised topological charge that is preserved under continuous deformations. Recent work on optical Stokes skyrmions in complex media has directly confirmed this robustness[8–10], showing that, provided boundary conditions are preserved, the associated topological charge remains highly stable even in the presence of realistic polarisation aberrations and noise. This inherent stability implies that topological indices such as the skyrmion number provide a discrete and robust label for an otherwise continuous optical field, motivating the use of skyrmions as information carriers for new forms of optical information processing[11,12].

An essential step toward skyrmion-based optical information processing is the generation of optical skyrmions, which has seen rapid advances. In the near field, surface plasmon polariton interference has produced skyrmion lattices that have been imaged with phase-resolved near-field microscopy[13]. In free space, tunable skyrmionic beams and related polarisation textures have also been engineered[14] via a variety of different techniques including liquid crystal droplets[15], spatial light modulators[3,16] and gradient index systems[17,18]. Additionally, metasurface-assisted surface-emitting lasers point to compact sources[19], and advances in meta-optics offer rich control of polarisation states along the optical path[20–22]. Lastly, q-plates, originally introduced for optical vortex generation, have also recently been extended to skyrmion generation[23,24].

Despite these promising advances, most studies remain focused on the generation of specific subtypes of optical skyrmions, such as fixed Néel-type[13] or Bloch-type[25,26] textures, and a broadly applicable framework for generation is still lacking. Moreover, in potential applications that exploit optical skyrmions as information carriers for optical computing and signal processing, generation alone is not sufficient: one must also be able to freely manipulate the generated textures to perform programmable operations on their topological charge. Lastly, current demonstrations of skyrmion manipulation typically rely on single-channel schemes, and establishing the practicality of multichannel operation remains an important goal.

Taken together, these considerations motivate the central question addressed in this paper: how can we devise a broadly applicable framework for the manipulation of polarisation textures, encompassing both their generation and a library of programmable logic operations, and ultimately extend these concepts to multichannel, parallel architectures? In this work, we adopt the notion of an arbitrary retarder[16]—here referring to spatially varying, tunable elliptical retarder based matter fields—to achieve this. Note that retarder fields are inherently device-agnostic, with a designed field capable of being implemented on multiple platforms including liquid crystal devices and metasurfaces[27–29] to realise the desired textures, establishing a flexible and broadly applicable framework for skyrmion manipulation. By cascading multiple retarder fields[30], this framework naturally accommodates both generation and logic, with each subsequent field acting on the textures produced upstream to execute required operations.

Building on these ideas, we present a platform-independent operational framework for manipulating topological polarisation textures based on cascades of retarder fields, as depicted in Fig. 1a,b. The framework adopts merons as the minimal logic units and consists of four essential optical primitives (Fig. 1c) each responsible for a specific conversion between topological textures: a generator which converts uniform fields (e.g. from light sources) to target polarisation textures, a skyrmion-meron interconverter, a meron-to-meron register, and a skyrmion-to-skyrmion adder. Together, these four optical primitives form a minimal operational toolkit for logic, whose roles are as follows:

(1) the generator synthesises skyrmions and merons with arbitrary topological charge;
(2) the skyrmion-meron interconverter (henceforth the converter) transforms skyrmions into merons and vice versa;
(3) the meron-to-meron register (henceforth the register) is devoted exclusively to transporting and storing merons without changing their boundary state and topological charge;
(4) the skyrmion-to-skyrmion adder (henceforth the adder) performs arithmetic operations on skyrmions.

With this basic set of primitives, skyrmion fields serve as computational states while meron fields serve as storage states. Apart from practical considerations relating to device complexity (see Results), this distinction between the roles played by skyrmions and merons serves as an important abstraction that simplifies logic design: a skyrmion is converted into a meron when it is not in use, and converted back into a skyrmion when required. However, note that this library can be further expanded, for example by incorporating meron-to-meron adders, to realise more advanced logical operations. Lastly, note that each primitive exhibits strong topological robustness and the underlying retarder field can be arbitrarily deformed without affecting its operation, provided that the boundary of the primitive remains unchanged (Methods 1 and 2).

To demonstrate the practical feasibility of our framework, we implement programmable retarder fields realised with liquid crystal spatial light modulators (LC-SLMs), and observe accurate charge readout under systematic and random errors (such as complex aberrations and noise[31–34]), consistent with our theoretical predictions. Beyond single-channel operation, we also realise multichannel, parallel implementations built from the same library of primitives. The work presented in this paper paves the way for the development of topologically robust programmable logic arrays using optical skyrmions.

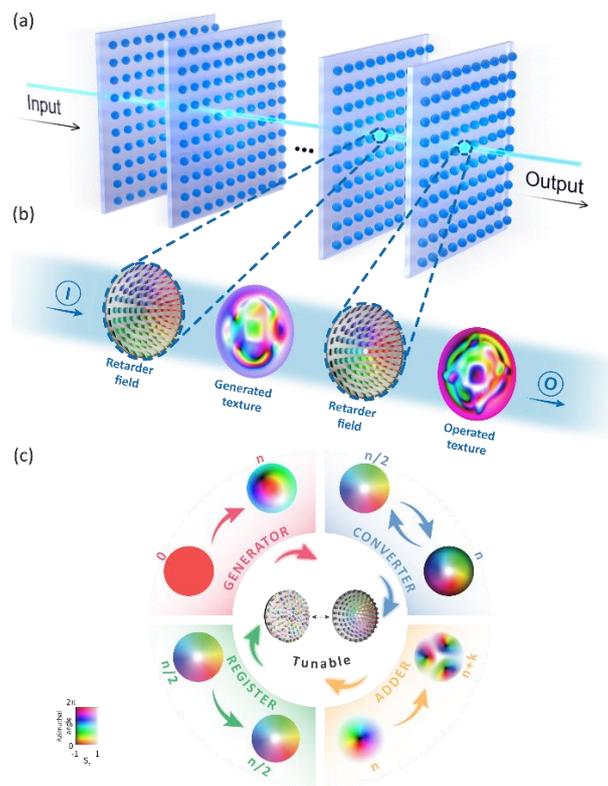

**Figure 1: Concept. (a)** A multichannel array capable of simultaneously performing a large number of distinct logic operations, thereby enabling practical skyrmion-based optical information processing. **(b)** Each unit of the array consists of a designed, spatially varying retarder selected from a library of distinct units. These units support both generation and logical operations, and can be cascaded to realise complex logic. Throughout this work, Stokes fields are depicted using colour to represent the azimuthal angle on the Poincaré sphere and saturation to represent height, whereas retarder fields are depicted using polarisation ellipses to indicate fast axis orientation. The input field is denoted by **I**, while the output field is denoted by **O**. **(c)** The four primitives introduced in this work, forming a minimal toolkit for skyrmion-based logic: generators, converters, registers, and adders.

## Results

As discussed in the Introduction, the key attribute that enables topological textures to function as information carriers is the presence of a stable, well-defined topological charge associated with the field. In the case of optical skyrmions, the relevant topological index is the integer-valued skyrmion number (see Supplementary note 1 and 2 for detailed definitions), derived from the usual notion of the degree of a map, and can be computed via the integral

$$N_{sk} = \frac{1}{4\pi} \iint S \cdot \left(\frac{\partial S}{\partial x} \times \frac{\partial S}{\partial y}\right) dxdy. \qquad (1)$$

Note, however, that integer quantisation of the above expression requires the field to satisfy specific boundary conditions, often referred to as compactifiability[8]. Without these boundary conditions, the equation above can no longer be identified with the degree of a map, but the field nonetheless exhibits topological character in the sense that the fractional part of $N_{sk}$ depends only on the boundary condition of the field[8].

Exploiting this property, we adopt both skyrmions and merons (for which the field is non-compactifiable and $N_{sk}$ lies in $\mathbb{Z}/2$) as basic units for information processing. Such an approach offers two key advantages. First, from the perspective of logic design, having two distinct units allows them to be assigned different roles, which in our case correspond to information processing (skyrmions) and information storage (merons). Second, increasing $N_{sk}$, whether integer or fractional, leads to increasing complexity of the underlying polarisation field. This, in turn, results in practical engineering limits due to limited resolution of the underlying matter field arising from finite pixel size. Adopting merons and skyrmions together thus overcomes these real-world issues as thrice as many distinct logic units are available for any given maximum topological charge, allowing the same operation to be realised with a third the maximum topological charge, and thereby reducing overall system complexity. Naturally, other fractional units could be employed to further increase the number of computational units, and we leave a detailed investigation of their design as a possible extension of this work.

The key approach for designing primitives that convert between different units relies on precise manipulation of the field boundary (Methods 1). As a result, the functionality of each unit depends only on how the incident field's boundary curve deforms as the field propagates through the device. This gives rise to strong topological robustness, similar to that established in Ref[11], whereby perturbations to the underlying retarder field that do not act on the boundary do not affect device functionality (Methods 2). In addition to relaxing fabrication and engineering constraints, this property is well suited to cascaded architectures, since the operation of each unit depends only on the boundary curve of the field exiting the preceding unit.

Adopting the approach outlined above, we begin by introducing the first of our four proposed primitives, namely generators which convert uniform incident light into skyrmions and merons of arbitrary order. The underlying retarder field features a fast axis distribution corresponding to an order-n meron, together with a radial retardance profile $\delta(r) = \pi r$ for skyrmion generation and $\delta(r) = \pi r/2$ for meron generation (Methods 1).

These generate skyrmions of order n and merons of order n/2, respectively.

To demonstrate the feasibility of the framework, we realise the retarder field using a cascade of three liquid-crystal spatial light modulators (LC-SLMs), as shown in Fig. 2a (Supplementary note 3 for experimental details and phase decomposition algorithm). In parallel, we characterise the retarder field via Mueller matrix polarimetry[31,35–40] and extract its fast axis orientation by Lu–Chipman decomposition (Supplementary note 4).

Fig. 2c presents experimentally generated Stokes fields[41,42] for various skyrmion generators designed using the framework above, together with the measured fast axis distributions of the corresponding primitives. Simulated results are also shown for comparison. Notice from the figure that since our primitives are designed to perform a specific operation on boundary curves, the resultant Stokes fields are not confined to any single subtype (Supplementary note 5). The numerically computed topological charge, evaluated using Eq. (1), is also shown, confirming that the primitives generate skyrmions with the correct topological charge. Lastly, Supplementary note 6 presents additional experimental results, with examples of meron generation.

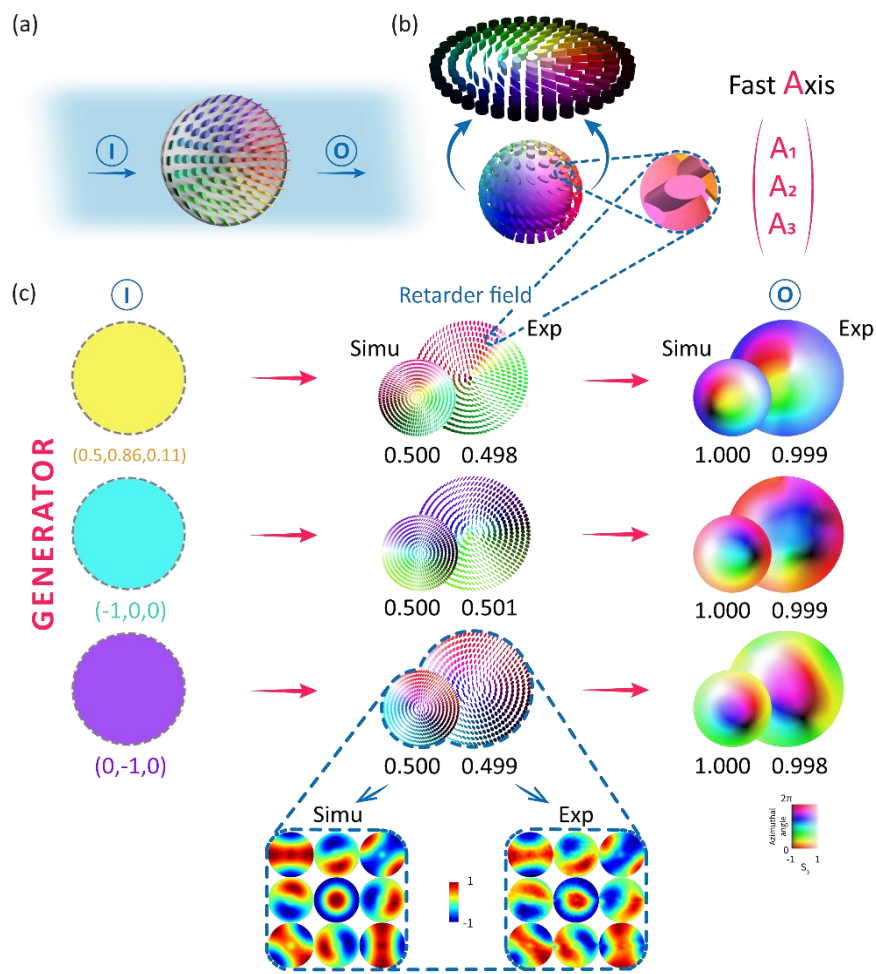

**Figure 2: Skyrmion generator. (a)** Input field passes through an arbitrary retarder field, and the resulting output field forms Stokes skyrmions. **(b)** A magnified view of a single element within the fast axis distribution is shown, which depicts the vector $\mathbf{A} = (A_1, A_2, A_3)^\mathrm{T}$ (Supplementary note 4). The colour of the ellipse represents the azimuthal orientation of the fast axis, while its ellipticity

represents the height. **(c)** Theoretical and experimentally measured Stokes fields generated by various cases. The input fields are uniform, while the output fields form skyrmions with a topological charge of 1. The retarder field of each case with a topological charge of 0.5 is also shown, depicted through the fast axis distribution obtained via Lu–Chipman decomposition[43]. Lastly, the simulated and experimentally measured Mueller matrix corresponding to the case of −45° polarised incident field is given.

Next, we elevate the platform to logic. To achieve this, we introduce a second retarder field downstream. In this stage, the generated Stokes fields are manipulated via three primitives: the converter, the register, and the adder (Fig. 1c). An important feature of our proposed design is the potential reusability of different matter fields. More specifically, in converters, the underlying retarder field is the same as that used for the meron generator, while in adders and registers it is the same as that used for the skyrmion generator (in the case of registers, the underlying retarder field exhibits a boundary fast-axis distribution aligned with the incident field and may adopt an arbitrary retardance profile; $\delta(r) = \pi r$ is used in this work for consistency). This demonstrates the multifunctionality of our approach, which is crucial for performing different computations using the same architecture, and underscores the modular nature of the designed optical primitives. In each case, the underlying matter field is similarly a meron-like field, with a detailed description presented in Methods 1.

With the primitives defined, Fig. 3a presents experimentally measured Stokes fields and corresponding simulations for representative examples. Specifically, results are shown for the converter, register, and adders of orders 2, 3, and 4, respectively. In each case, the fast axis distributions extracted from the Mueller matrices are also presented. The agreement is excellent, as the measured Stokes fields closely match the simulations and the charge readout agrees with the design within experimental uncertainty. Importantly, note that while only a single operation is demonstrated here, the primitives introduced can be further cascaded to achieve complex logic operations (see Method 3).

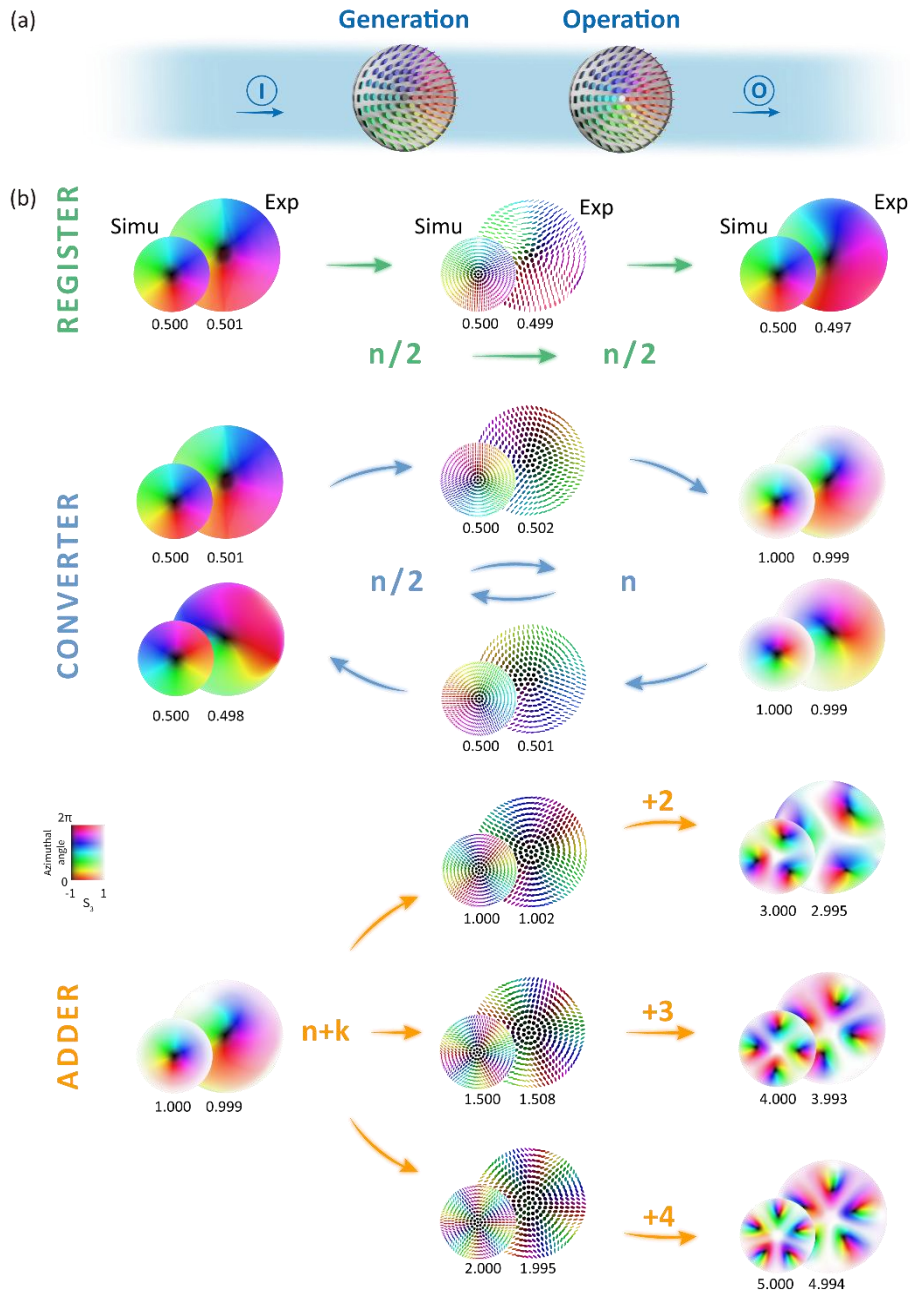

**Figure 3: Primitives for Logic. (a)** Schematic illustrating the cascaded architecture used to perform logic operations sequentially. **(b)** Simulated and experimentally measured Stokes fields of the input and output fields for each primitive. The corresponding fast axis distribution for each primitive is also shown. (Top) A register that converts an order-1 meron into another order-1 meron. (Middle) A converter which switches between an order-1 meron and an order-2 skyrmion. (Bottom) Adders of orders 2, 3 and 4 acting on a skyrmion of order-1.

Here, to demonstrate parallel operation and architectural reusability, we implement a multichannel array utilising a dual-stage configuration (Fig. 4a). More specifically, we employ a reconfigurable layer for state preparation followed by a fixed processing layer for operation. Here, we focus on a representative $3 \times 3$ array in which the various operations which change according to input can be clearly observed (see Methods 3 for a higher-order 10 × 10 implementation).

As shown in Figure 4b, within the annotated bottom-right 2 × 2 area of the array, different functionalities including skyrmion and meron generation ($G_S$ and $G_M$), as well as information relay (R), conversion (C), and addition (A) are achieved when different input fields are changed. Moreover accurate skyrmion numbers derived from the experimental Stokes fields (bottom panel) not only validate the execution of these operations but also serve as quantitative confirmation of independent channel configuration, demonstrating negligible crosstalk and high robustness against complex aberrations and noise. Crucially, these results show that a single reconfigurable stage (in comparison to the two reconfigurable stages used in previous demonstrations) can be used to perform varied and complex operations, offering significant real-world advantages, as reconfigurable components are often larger, more costly and difficult to fabricate.

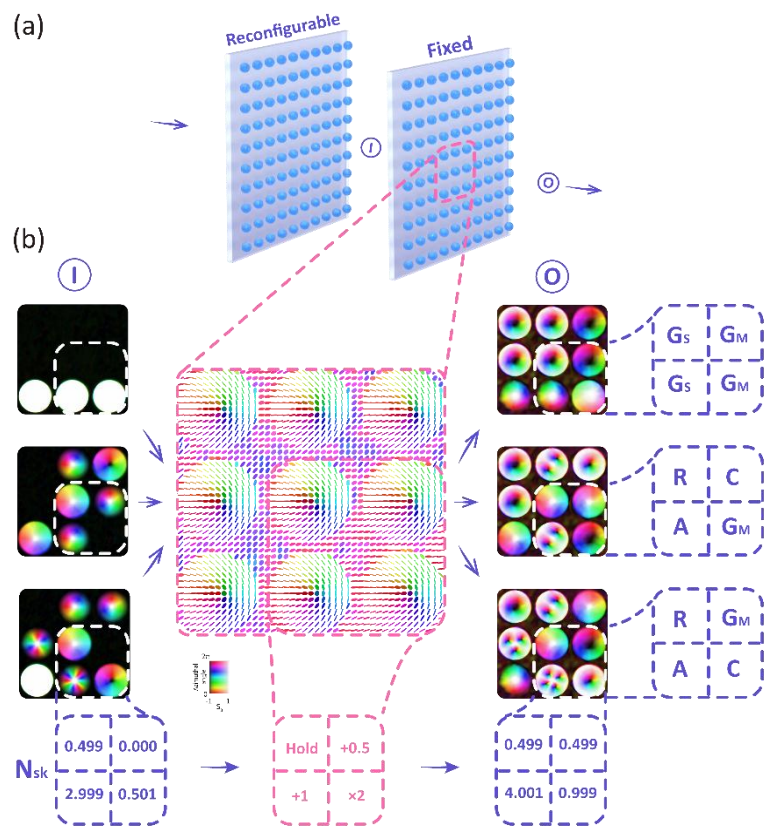

**Figure 4: Multichannel array and functional reusability. (a)** A reconfigurable stage for state preparation followed by a fixed processing stage to demonstrate architectural reusability. **(b)** A multichannel array is used to demonstrate the simultaneous performance of distinct logic operations and the reusability of the architecture. Three different sets of 3 × 3 incident arrays passed through the same 3 × 3 retarder field array to produce various output configurations, illustrating that a single reconfigurable platform can be repurposed for multiple functions by modulating the input state. The specific operations performed by the bottom-right 2 × 2 area are annotated: $G_S$ (skyrmion generator), $G_M$ (meron generator), R (register), C (converter), and A (adder). For the third (bottom) set of results, the experimentally measured skyrmion numbers $N_{sk}$ are provided alongside their corresponding arithmetic logic.

## Discussion

In this work, we present and experimentally validate a platform-independent operational framework for manipulating the topological charge of polarisation textures, enabling skyrmion-based optical processing encompassing generation, conversion, storage, and arithmetic. Our approach uses a library of pre-designed topologically robust optical primitives, which modularises the design of logic and enables systematic composition, reuse, and scaling of complex operations. An important aspect of our proposed approach is that various different operations can be realised with the same underlying primitive. This enables a single reconfigurable generation layer to act as a way of selecting between different operations implemented with a fixed processing layer, thereby achieving logic using only a single dynamic stage. As such, we anticipate that the framework presented in this paper will provide a practical foundation for scalable, reconfigurable skyrmion-based optical information processing.

From an engineering perspective, the topological nature of our approach implies strong robustness against systematic and random errors, a key property that is not only highly relevant for optical communications and interconnections[5], but also not present in existing optical computing strategies[44–46]. Indeed, our experiments confirm that the logic operations remain reliable despite practical imperfections, including misalignment between cascaded devices, retardance and angular errors, the temperature dependence of liquid crystal devices and ambient mechanical vibrations. This topological robustness relaxes manufacturing constraints, suggesting that our framework is not limited to LC-SLM cascades but can be physically realised using diverse technologies, including passive liquid crystal polymers[47,48], laser-written birefringent glasses[49], or nanophotonic metasurfaces[50,51]. Such platform flexibility means that the approach demonstrated here can potentially be transferred from a proof-of-concept to more compact and integrated photonic logic circuits.

Before concluding, several important extensions of this work are worth discussing. First, while this work considers free-space propagation between layers, real on-chip systems relay information via confined media, such as waveguides, for which topological preservation is not yet fully understood. As such, a systematic investigation of topological preservation under waveguide confinement and modal dispersion constitutes an important extension of this work[5]. Second, while we consider only the conventional skyrmion number in this work, it is possible to extend the adder operation to the generalized skyrmion number[52], which has the potential to further enhance data density. Third, while topological protection provides strong robustness against fabrication errors within each channel, the quality of the boundary polarization states remains an important challenge. This motivates the development of cutting-edge vectorial adaptive optics strategies that focus on compensating such errors[33,53].

To conclude, despite growing interest in photonic computing as a low-power, high-bandwidth approach for specialised information processing[45,46,54–56], robustness to noise remains a central challenge for scalability[44–46]. Optical skyrmions address this challenge by providing topologically protected information carriers that are intrinsically resilient to noise; however, the novelty of this approach means that high-level architectural design remains an open challenge. In this paper, we introduce a set of units that form the building blocks of a higher-level architectural framework for skyrmion-based optical information processing, providing a concrete starting point for system-level design. As such, we believe these primitives offer a practical route toward scalable photonic architectures that exploit topological robustness for reliable information processing.

## Methods

### 1 Designing primitives

In this section, we describe the key steps for designing primitives to manipulate the skyrmion number integral in a topologically robust way. Here, we consider fields defined on the closed unit disc, which we denote by $\Omega$. The central idea relies on the results proven in Ref[8,52], which demonstrate that the boundary condition of a field determines the fractional part of the skyrmion number integral. Using curly brackets $\{\cdot\}$ to denote the fractional part of a real number, this fact implies the existence of a well-defined map $\mathcal{T}: C^\infty(\partial\Omega, S^2) \to S^1$ given by

$$\mathcal{T}(\Sigma) = e^{2\pi i \{N_{sk}(S)\}} \tag{2}$$

where $S: \Omega \to S^2$ is any smooth field satisfying $S|_{\partial\Omega} = \Sigma$, and $N_{sk}(S)$ the corresponding skyrmion number integral as given by Eq. (1).

Consider now a retarder field whose fast axis distribution is given by an order-n meron with a radial retardance profile $\delta(r)$. Setting $\sigma = \delta(1)$, if the boundary of the incident field is uniformly right-circularly polarised, the boundary of the output field can be computed directly as

$$\Sigma_\sigma(\theta) = \begin{pmatrix} \sin\sigma \sin n\theta \\ -\sin\sigma \cos n\theta \\ \cos\sigma \end{pmatrix}, \tag{3}$$

from which one derives

$$\mathcal{T}(\Sigma_\sigma) = e^{\pi i \{n(1-\cos\sigma)\}}. \tag{4}$$

Now, let $H: \Omega \times [0,1] \to S^2$ be the homotopy induced by propagation through the retarder. More specifically, the field at $t \in [0,1]$ is given by the action of a retarder field with fast axis distribution corresponding to an order-n meron and a retardance profile of $t\delta(r)$. One easily verifies that $\mathcal{T} \circ H|_{\partial\Omega}: [0,1] \to S^1$ is given by

$$\mathcal{T} \circ H|_{\partial\Omega}(t) = e^{\pi i \{n(1-\cos t\sigma)\}}. \tag{5}$$

But by definition, we have

$$\mathcal{T} \circ H|_{\partial\Omega}(t) = e^{2\pi i \{N_{sk}(H|_{\partial\Omega}(t))\}}, \tag{6}$$

and $N_{sk}(H|_{\partial\Omega}(t))$ is continuous with respect to t. Thus, $N_{sk}(H|_{\partial\Omega}(1))$ is given by the unique lift of $e^{\pi i\{n(1-\cos t\sigma)\}}$ starting at $N_{sk}(H|_{\partial\Omega}(0))$. This implies that the difference in skyrmion numbers before and after the retarder field is a fixed quantity

$$\Delta N_{sk} = \int_0^1 \mathcal{T} \circ H|_{\partial\Omega}^* \, d\theta \tag{7}$$

where $d\theta$ is the standard angular 1-form on $S^1$. By direct computation, this implies that if the boundary of the incident field is uniformly right-circularly polarised, then

1. $\Delta N_{sk} = 0.5n$ if $\sigma = \pi/2$ and hence behave as meron generators and skyrmion to meron converters.

2. $\Delta N_{sk} = 0.5n$ if $\sigma = \pi$ and hence behave as skyrmion generators and skyrmion-skyrmion adders.

Repeating this argument for the case where the boundary of the incident field is itself a meron (Supplementary note 1), one has that when $\sigma = \pi/2$, the retarder field behaves as a meron to skyrmion converter. Lastly, in the case of registers, we consider an incident field whose boundary corresponds to a meron. If the fast axis distribution at the boundary of the underlying retarder field aligns with the polarisation state of the incident field, propagation through the system leaves the boundary conditions unchanged, and the skyrmion number is therefore preserved.

Two important remarks are in order. First, the derivation above demonstrates that the change in skyrmion number induced by a retarder field depends only on how the boundary curve of the incident field transforms during propagation. This, in turn, depends only on the structure of the retarder field at its boundary. Together, these facts imply that, as in Ref[11], any perturbation to the underlying retarder field that does not affect the boundary has no impact on the functionality of the primitives, which therefore exhibit strong topological robustness.

Second, in our analysis above, we consider only the case in which the incident boundary field is uniformly right-circularly polarised. However, only minor modifications are required to generalise this to other incident fields. In particular, if the boundary of the incident field is in any other polarisation state, let $G$ be a rotation matrix that maps right-circularly polarised light to that state and $R$ be any relevant matter field described above. Then the primitive given by $GRG^T$ achieves the same functionality for the altered incident boundary state (Supplementary note 5 for details).

## 2 Topological protection of primitives

To explicitly validate the resilience of the proposed logic architecture against environmental noise and fabrication imperfections, we experimentally characterised the stability of the topological charge under controlled perturbations, as presented in Fig. 5. Using the skyrmion generator as a reference case, we introduced random phase fluctuations onto the retarder field. The disorder profiles were generated with a spatial frequency cutoff of $K_{max} = 0.05$ (normalised units) to mimic spatially correlated errors, and scaled by a maximum retardance $\Phi_{max}$ ranging from 0 to 20 rad. Crucially, to rigorously test the topological protection mechanism defined by the boundary condition, a radial envelope proportional to $(1 - r^2)$ was applied to the noise map. This ensured that while the internal texture of the device was severely scrambled, the boundary condition remained unperturbed. The resulting distorted Stokes fields were measured and are presented in Fig. 5b. Despite significant optical distortion visible in the intensity profiles, the integrated skyrmion number remains invariant at $N_{sk} \approx 1$ across all noise levels (Fig. 5c). This result confirms that the topological protection mechanism ensures reliable operation even in the presence of substantial disorder, validating the robustness claims discussed in the main text.

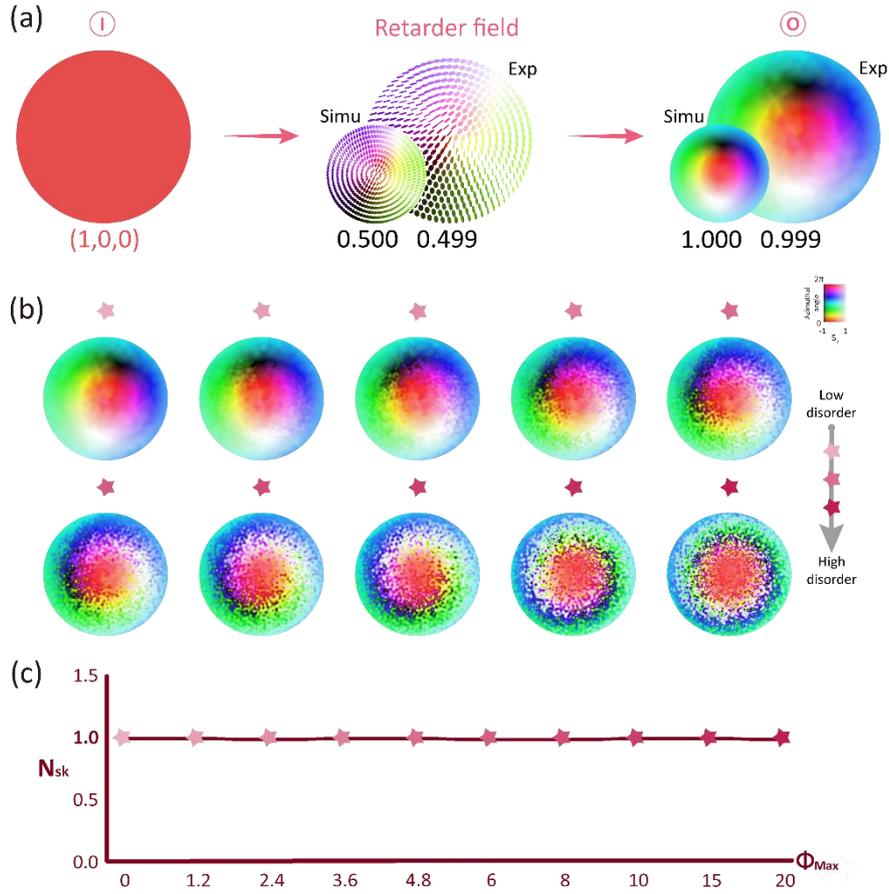

**Figure 5: Topological robustness to noise. (a)** Reference case: Generation of an $N_{sk} = 1$ skyrmion from a horizontally polarised incident state. **(b)** Output results under ten increasing levels of disorder from 0 to 20 rad. Star markers indicate the noise level, with darker colours representing higher levels of perturbation. **(c)** Calculated skyrmion number $N_{sk}$ plotted against the noise level $\Phi_{max}$.

## 3 Large-scale demonstration

Lastly, to demonstrate the scalability of the proposed framework beyond the representative 3 × 3 array shown in the main text (Fig. 4), we extended the architecture to a 10 × 10 array of independent channels. This large-scale implementation utilises the same library of topological primitives to perform parallel generation and manipulation on 100 spatially multiplexed beams. To illustrate the functional diversity, four specific channels (circled in Fig. 6) are highlighted; from top to bottom, these channels correspond to generator, register, register, and adder operations, respectively, with their resulting skyrmion numbers explicitly indicated. The experimentally measured Stokes fields confirm that the framework maintains high performance characterised by accurate charge readout and negligible crosstalk even when scaled to significantly higher channel densities, validating that the intrinsic topological robustness persists in large-scale implementations. These results confirm that scaling to larger arrays offers a natural route to parallelism and increased computational power.

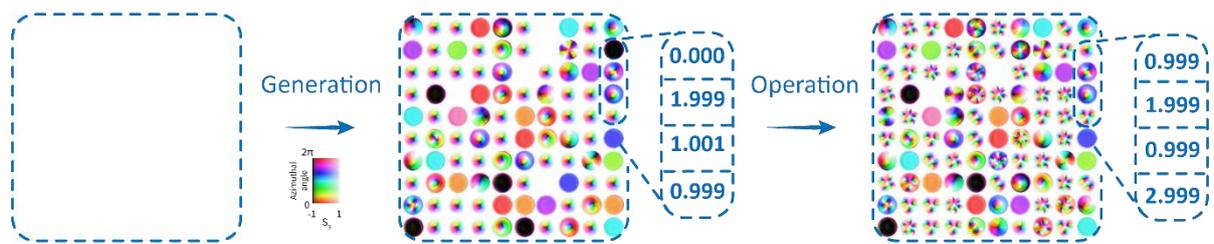

**Figure 6: Multichannel array.** Experimentally measured Stokes fields of a $10 \times 10$ array of independent channels. Each channel is modulated by a different primitive in two stages, enabling rapid generation of 100 beams followed by subsequent operations on their skyrmion numbers. The skyrmion number of selected output units are also shown, where accurate charge read-out is demonstrated.

# References

1. Shen, Y. *et al.* Optical skyrmions and other topological quasiparticles of light. *Nat. Photonics* **18**, 15–25 (2024).

2. Zhang, Y. *et al.* Skyrmions based on optical anisotropy for topological encoding. (2025).

3. Wu, H., Zhou, W., Zhu, Z. & Shen, Y. Optical skyrmion lattices accelerating in a free-space mode. *APL Photonics* **10**, (2025).

4. Ma, Y. *et al.* Using optical skyrmions to assess vectorial adaptive optics capabilities in the presence of complex aberrations. *Sci. Adv.* **11**, (2025).

5. Wang, A. A. *et al.* Optical Skyrmions in Waveguides. (2025).

6. Yessenov, M. *et al.* Ultrafast space-time optical merons in momentum-energy space. *Nat. Commun.* **16**, 8592 (2025).

7. Król, M. *et al.* Observation of second-order meron polarization textures in optical microcavities. *Optica* **8**, 255 (2021).

8. Wang, A. A. *et al.* Topological protection of optical skyrmions through complex media. *Light Sci. Appl.* **13**, 314 (2024).

9. Peters, C., Everts, K., Kleine, T., Ornelas, P. & Forbes, A. Seeing through randomness with topological light. (2025).

10. Bezuidenhout, H., Peters, C., Kumar, R., Forbes, A. & Nape, I. Deep diffractive optical neural networks for detecting Skyrmionic topologies of light. (2025).

11. Wang, A. A. *et al.* Perturbation-resilient integer arithmetic using optical skyrmions. *Nat. Photonics* **19**, 1367–1375 (2025).

12. He, C., Shen, Y. & Forbes, A. Towards higher-dimensional structured light. *Light Sci. Appl.* **11**, 205 (2022).

13. Tsesses, S. *et al.* Optical skyrmion lattice in evanescent electromagnetic fields. *Science (1979)*. **361**, 993–996 (2018).

14. Zhu, J., Liu, S., Zhang, Y.-S., Li, C.-F. & Guo, G.-C. Synthesis and observation of optical skyrmionic structure in free space. *Phys. Rev. A (Coll Park)*. **110**, 043522 (2024).

15. Kamal, W. *et al.* On-demand pitch tuning of printed chiral nematic liquid crystal droplets. *Mater. Today Adv.* **19**, 100416 (2023).

16. He, C. *et al.* A reconfigurable arbitrary retarder array as complex structured matter. *Nat. Commun.* **16**, 4902 (2025).

17. He, C. *et al.* Complex vectorial optics through gradient index lens cascades. *Nat. Commun.* **10**, 4264 (2019).

18. Shen, Y. *et al.* Topologically controlled multiskyrmions in photonic gradient-index lenses. *Phys. Rev. Appl.* **21**, 024025 (2024).

19. Shitrit, N. Surface-emitting lasers meet metasurfaces. *Light Sci. Appl.* **13**, 37 (2024).

20. Dorrah, A. H., Rubin, N. A., Zaidi, A., Tamagnone, M. & Capasso, F. Metasurface optics for on-demand polarization transformations along the optical path. *Nat. Photonics* **15**, 287–296 (2021).

21. Li, X., Lan, T.-H., Tien, C.-H. & Gu, M. Three-dimensional orientation-unlimited polarization encryption by a single optically configured vectorial beam. *Nat. Commun.* **3**, 998 (2012).

22. Gu, T., Kim, H. J., Rivero-Baleine, C. & Hu, J. Reconfigurable metasurfaces towards commercial success. *Nat. Photonics* **17**, 48–58 (2023).

23. Hakobyan, V., Shen, Y. & Brasselet, E. Unitary spin-orbit optical-skyrmionic wave plates. *Phys. Rev. Appl.* **22**, 054038 (2024).

24. Hakobyan, V. & Brasselet, E. Q-Plates: From Optical Vortices to Optical Skyrmions. *Phys. Rev. Lett.* **134**, 083802 (2025).

25. Lei, X. *et al.* Photonic Spin Lattices: Symmetry Constraints for Skyrmion and Meron Topologies. *Phys. Rev. Lett.* **127**, 237403 (2021).

26. Gao, S. *et al.* Paraxial skyrmionic beams. *Phys. Rev. A (Coll Park).* **102**, 053513 (2020).

27. Rubin, N. A., Shi, Z. & Capasso, F. Polarization in diffractive optics and metasurfaces. *Adv. Opt. Photonics* **13**, 836 (2021).

28. Shaltout, A. M., Shalaev, V. M. & Brongersma, M. L. Spatiotemporal light control with active metasurfaces. *Science (1979).* **364**, (2019).

29. Dorrah, A. H. & Capasso, F. Tunable structured light with flat optics. *Science (1979).* **376**, (2022).

30. Wang, A. A., Cai, Y., Ma, Y., Salter, P. S. & He, C. Resolving topological obstructions to vectorial structured field control. (2026).

31. Lin, J. *et al.* Full Poincaré polarimetry enabled through physical inference. *Optica* **9**, 1109 (2022).

32. Wang, A. A. *et al.* General in situ feedback control of cascaded liquid crystal spatial light modulators for structured field generation. (2026).

33. He, C., Antonello, J. & Booth, M. J. Vectorial adaptive optics. *eLight* **3**, 23 (2023).

34. Shen, Y. *et al.* Polarization Aberrations in High-Numerical-Aperture Lens Systems and Their Effects on Vectorial-Information Sensing. *Remote Sens. (Basel).* **14**, 1932 (2022).

35. Azzam, R. M. A. Photopolarimetric measurement of the Mueller matrix by Fourier analysis of a single detected signal. *Opt. Lett.* **2**, 148 (1978).

36. Goldstein, D. H. Mueller matrix dual-rotating retarder polarimeter. *Appl. Opt.* **31**, 6676 (1992).

37. Azzam, R. M. A. Stokes-vector and Mueller-matrix polarimetry [Invited]. *Journal of the Optical Society of America A* **33**, 1396 (2016).

38. Zhang, Z. *et al.* Analysis and optimization of aberration induced by oblique incidence for in-vivo tissue polarimetry. *Opt. Lett.* **48**, 6136 (2023).

39. Deng, L. *et al.* Influence of hematoxylin and eosin staining on linear birefringence measurement of fibrous tissue structures in polarization microscopy. *J. Biomed. Opt.* **28**, (2023).

40. Deng, L. *et al.* A Dual-Modality Imaging Method Based on Polarimetry and Second Harmonic Generation for Characterization and Evaluation of Skin Tissue Structures. *Int. J. Mol. Sci.* **24**, 4206 (2023).

41. Cai, Y., Wang, A. A. & He, C. Rethinking conditioning in polarimetry: a new framework beyond $\ell^2$-based metrics. (2025).

42. He, C. *et al.* Polarisation optics for biomedical and clinical applications: a review. *Light Sci. Appl.* **10**, 194 (2021).

43. Lu, S.-Y. & Chipman, R. A. Interpretation of Mueller matrices based on polar decomposition. *Journal of the Optical Society of America A* **13**, 1106 (1996).

44. Zhou, W. *et al.* In-memory photonic dot-product engine with electrically programmable weight banks. *Nat. Commun.* **14**, 2887 (2023).

45. Feldmann, J. *et al.* Parallel convolutional processing using an integrated photonic tensor core. *Nature* **589**, 52–58 (2021).

46. Shen, Y. *et al.* Deep learning with coherent nanophotonic circuits. *Nat. Photonics* **11**, 441–446 (2017).

47. Zeng, H. *et al.* High-Resolution 3D Direct Laser Writing for Liquid-Crystalline Elastomer Microstructures. *Advanced Materials* **26**, 2319–2322 (2014).

48. Tartan, C. C. *et al.* Read on Demand Images in Laser-Written Polymerizable Liquid Crystal Devices. *Adv. Opt. Mater.* **6**, (2018).

49. Fedotov, S. S. *et al.* Direct writing of birefringent elements by ultrafast laser nanostructuring in multicomponent glass. *Appl. Phys. Lett.* **108**, (2016).

50. Devlin, R. C., Khorasaninejad, M., Chen, W. T., Oh, J. & Capasso, F. Broadband high-efficiency dielectric metasurfaces for the visible spectrum. *Proceedings of the National Academy of Sciences* **113**, 10473–10478 (2016).

51. Yu, N. *et al.* Light Propagation with Phase Discontinuities: Generalized Laws of Reflection and Refraction. *Science (1979).* **334**, 333–337 (2011).

52. Wang, A. A. *et al.* Generalized Skyrmions. (2024).

53. He, C. & Booth, M. J. Vectorial adaptive optics: correction of polarization and phase. in *Imaging and Applied Optics Congress 2022 (3D, AOA, COSI, ISA, pcAOP)* OTh3B.4 (Optica Publishing Group, Washington, D.C., 2022). doi:10.1364/AOA.2022.OTh3B.4.

54. Hua, S. *et al.* An integrated large-scale photonic accelerator with ultralow latency. *Nature* **640**, 361–367 (2025).

55. Ahmed, S. R. *et al.* Universal photonic artificial intelligence acceleration. *Nature* **640**, 368–374 (2025).

56. Sotirova, A. S. *et al.* Low cross-talk optical addressing of trapped-ion qubits using a novel integrated photonic chip. *Light Sci. Appl.* **13**, 199 (2024).